\documentclass[fleqn,usenatbib]{mnras}

\usepackage{newtxtext,newtxmath}

\usepackage[T1]{fontenc}

\DeclareRobustCommand{\VAN}[3]{#2}
\let\VANthebibliography\thebibliography
\def\thebibliography{\DeclareRobustCommand{\VAN}[3]{##3}\VANthebibliography}

\usepackage{graphicx}	
\usepackage{amsmath}	
\usepackage{caption}
\usepackage{subcaption}
\usepackage[utf8]{inputenc}
\usepackage{xcolor}

\newcommand{\Ssfr}        {\mbox{$\Sigma_{\rm SFR}$}}
\newcommand{\Sstar}       {\mbox{$\Sigma_{\star}$}}
\newcommand{\Smol}       {\mbox{$\Sigma_{\rm mol}$}}

\newcommand{\Mstar}      {\mbox{$M_\mathrm{{\star} }$}}

\newcommand{\fg}      {\mbox{$\mathrm{f_{mol} }$}}
\newcommand{\Dsfms}      {\mbox{$\mathrm{\Delta SFMS}$}}
\newcommand{\Dsk}      {\mbox{$\mathrm{\Delta SK}$}}

\newcommand{\Dmgms}      {\mbox{$\mathrm{\Delta MGMS}$}}
\newcommand{\Ha}          {\mbox{$\mathrm{H\alpha}$}}
\newcommand{\Hb}          {\mbox{$\mathrm{H\beta}$}}
\newcommand{\Reff}          {\mbox{$R_{\mathrm{eff}}$}}
\newcommand{\Msun}          {\mbox{$\mathrm{M_{\odot}}$}}
\newcommand{\CO}          {\mbox{$\mathrm{^{12}CO}$}}
\newcommand{\COIII}          {\mbox{$\mathrm{^{13}CO}$}}

\title[Radial profiles of galaxies in merger]{The EDGE-CALIFA Survey: Influence of Mergers on Radial Profiles of Star-Formation Properties}

\author[Y. Garay-Solis et al.]{
Y. Garay-Solis,$^{1}$ \thanks{E-mail: ygaray@astro.unam.mx}
J. K. Barrera-Ballesteros,$^{1}$
L. Carigi,$^{1}$
D. Colombo,$^{2}$
S. F. Sánchez,$^{3}$
\newauthor A. Z. Lugo-Aranda,$^{3}$
V. Villanueva,$^{4}$
T. Wong$^{5}$
and A. D. Bolatto$^{6}$
\\
$^{1}$Instituto de Astronom\'ia, Universidad Nacional Aut\'onoma de M\'exico, A.P. 70-264, 04510 CDMX, M\'exico\\
$^{2}$Argelander-Institut für Astronomie, Auf dem Hügel 71, 53121 Bonn, Germany\\
$^{3}$Instituto de Astronom\'ia, Universidad Nacional Aut\'onoma de M\'exico, A.P. 106, Ensenada, 22800, BC, M\'exico\\
$^{4}$Departamento de Astronom{\'i}a, Universidad de Concepci{\'o}n, Barrio Universitario, Concepci{\'o}n, Chile\\
$^{5}$Department of Astronomy, University of Illinois, Urbana, IL 61801, USA\\
$^{6}$Department of Astronomy, University of Maryland, College Park, MD 20742, USA
}

\date{Accepted XXX. Received YYY; in original form ZZZ}

\pubyear{\the\year{}}

\begin{document}
\label{firstpage}
\pagerange{\pageref{firstpage}--\pageref{lastpage}}
\maketitle

\begin{abstract}

In this study, we investigate how the merging process influences the radial variations of the specific Star Formation Rate (sSFR), Star Formation Efficiency (SFE), and molecular gas fraction (\fg) in galaxies. We analyse 33 isolated galaxies and 34 galaxies in four different merger stages from pairs, merging galaxies, post-mergers, and merger remnants. Our sample is included in the EDGE-CALIFA survey, which provides spatially resolved optical integral-field unit and CO spectroscopy data. We show that, in comparison with the isolated sample, the mergers increase the molecular gas fraction non-uniformly across different galactocentric distances. Also, we find that the main driver (efficiency or molecular gas) of both enhanced and suppressed star formation changes independently of galactocentric radius and merger stage. However, efficiency appears to be the primary driver of variations in star formation (except during the merging stage), where we find an enhancement in star formation driven by the available fuel. Our results suggest that in interacting and merging galaxies, the efficiency plays a crucial role in the star formation variations throughout the galaxy, regardless of the available molecular gas content.

\end{abstract}



\section{Introduction}
\label{sec:intro}

Galaxy interactions and mergers are known as crucial drivers of galaxy evolution, which can significantly alter their internal properties.
Early observations \citep[e.g.,][]{Arp1966,Larson1978} and numerical models \citep[e.g.,][]{Toomre1972,Hernquist1989} established the connection between tidal distortions, starbursts, and quasars, leading to the recognition of mergers as a fundamental process in galaxy formation within the $\Lambda$CDM framework \citep[e.g.,][]{Blumenthal1984,White1991}.

Subsequently, large imaging surveys coupled with single-fibre spectroscopy have provided a deeper understanding of the evolution of nearby galaxies \citep[e.g., Sloan Digital Sky Survey (SDSS),][]{SDSS_2000}. 
For instance, observational studies and simulations have shown that galaxy interactions significantly affect star formation, gas content, morphology, and nuclear activity, with the strength of these effects depending on pair separation, mass ratio, gas fraction, and environment \citep[e.g.,][]{Ellison2008,Ellison2010,Ellison2011,Patton2011,Scudder2012,Ellison2013,Patton2013,Satyapal2014,Scudder2015, Patton2016}.  

More recently, the development of integral field units (IFUs) has advanced these studies through the first generation of optical integral field spectroscopy (IFS) galaxy surveys, e.g., CALIFA \citep[Calar Alto Legacy Integral Field Area;][]{califa2012} survey and MaNGA \citep[Mapping Nearby Galaxies at Apache Point Observatory;][]{Manga} survey, allowing for the exploration of galaxies at kpc-scales.
For instance, \citet{Pan2019} found that galactic interactions initially exert minimal influence on radial specific Star Formation Rate (sSFR) profiles. However, after the first pericentric passage, the radial profile drops steeply from enhanced to suppressed sSFR for increasing galactocentric radii. In later merger stages, sSFR is enhanced over a wide spatial scale up to $\sim6.7$ kpc, and the enhancement is generally peaked at the centre \citep{Thorp2019,Pan2019,Steffen2021}. In the outer regions, \citet{Steffen2021} reported weak suppression of sSFR $\sim0.1$ dex, whereas \citet{Pan2019} suggest that star formation enhancement is not restricted only to the central region of galaxies. Also, \citet{Thorp2019} found enhancements and suppressions in the Star Formation Rate (SFR) in the galactic outskirts.
A recent study by \citet{Thorp_2024} compares spatially resolved SFR profiles in post-merger galaxies with non-interacting galaxies that exhibit similar global star formation enhancements. They found that post-merger galaxies exhibit a much stronger central starburst than isolated galaxies. 
On the other hand, studies that include observational CO data analysed three galaxies with recent merger signals find that star formation is primarily driven by the efficiency with which molecular gas is converted, which is similar to isolated galaxies \citep{Ellison2020}.
Nevertheless, according to \citet{Thorp2022}, merger-induced star formation can be driven by both the efficiency and the amount of available molecular gas, and this is independent of whether the galaxy is in a merger stage or isolated.
Therefore, to reliably quantify the role of the merger process in star formation behaviour, studies that combine optical and submillimeter observations at kpc-scale are required. 

Although pioneering IFU surveys have provided unprecedented detail, there is still no clear consensus on what drives merger-induced star formation and where within the galaxy it predominates. It is still debated whether the observed changes are driven primarily by gas (fuel) accretion or by local variations in star formation efficiency, and how this changes along the merger sequence. In this work, we address these questions by performing a radial analysis of the specific star formation rate (sSFR), star formation efficiency (SFE), and molecular gas fraction (\fg) of galaxies at different merger stages. Our goal is to determine how these properties are affected during a merger event in different galactic regions,  thereby building a more detailed picture of how mergers redistribute gas and impact the drivers of star formation across different galactic regions. We explore a sample of 33 isolated galaxies and 34 galaxies at four merger stages. Our galaxies have CO data from the Extragalactic Database for Galaxy Evolution (EDGE) survey \citep{Bolatto2017,Wong2024}, and optical IFS data from the CALIFA survey \citep{califa2012}, covering galactocentric distances up to $\sim2.5$ effective radius (\Reff). The EDGE-CALIFA sample provides an ideal laboratory to explore how these physical properties change during merger events compared to those observed in non-interacting star-forming galaxies.

This paper is organised as follows: in Section~\ref{sec:data_analysis} we describe the key characteristics of the data and sample; in Section~\ref{sec:Results} we present the analysis results for each merger stage compared to the control sample; in Section~\ref{sec:discusion} we discuss our findings. Finally, in Section~\ref{sec:conclusions} we summarize our analysis and present the main results of our study.
For the derived data, we adopt a cosmology with the following parameters:
$\mathrm{H_0=70\,km\,s^{-1}\,Mpc^{-1}}$, $\mathrm{\Omega_M=0.27}$, and $\mathrm{\Omega_\Lambda=0.73}$.

\section{Data, Sample, and Analysis} 
\label{sec:data_analysis}

\subsection{The EDGE-CALIFA survey}
\label{subsec:EDGE_CALIFA}

We analyse a sample of galaxies that combine the CO spectroscopy from the EDGE survey \citep{Bolatto2017,Wong2024} with the optical IFS data from the CALIFA survey \citep{califa2012},  referred to as the EDGE-CALIFA survey. For detailed information on the data and their calibration, we refer readers to \citet{Bolatto2017}, \citet{Wong2024}, and \citet{califa2012}. In the next paragraphs, we briefly overview the main features of these datasets.

The CALIFA survey observed more than 600 galaxies in the local universe with redshifts of\, $0.005<z<0.03$.
The PPAK Integral Field Unit (IFU) \citep{PPAK2006} was attached to the 3.5~m telescope at the Calar Alto Observatory, providing a 74~$\times$~64 arcsec Field-of-View (FoV) and covering $\sim2.5~\Reff$ with a physical spatial resolution of $\sim$1~kpc (FWHM$\sim$2.5 arcsec) for each galaxy.
The CALIFA survey used the Potsdam Multi Aperture Spectrograph (PMAS, \citealt{PMAS2005}) with two configurations: a low spectral resolution of 6~\AA\ (V500; 3750-7000~\AA) and an intermediate spectral resolution of 2.7~\AA\ (V1200; 3700-4700~\AA). The reduction of the PMAS/PPAK data was performed using a specialised pipeline developed for the CALIFA survey \citep{Califa_reduccion}. The physical properties for each galaxy were extracted by analysing the stellar populations and emission lines of individual spaxels using the PIPE3D \citep{pipe3D1,pipe3D2}.

The EDGE survey observed a subsample of 125 galaxies from the CALIFA survey, using the Combined Array for Research in Millimeter-wave Astronomy \citep[CARMA,][]{carma2006}. The EDGE survey complements the optical data from the CALIFA survey by providing \CO(1-0) and \COIII(1-0) maps with the same FoV coverage as the CALIFA.
The \CO(1-0)  maps combine data from both E and D configurations of CARMA, reaching a physical resolution $\sim$1.4~kpc (FWHM$\sim$4.5 arcsec) \citep{Bolatto2017}.

The Python-based EDGE-CALIFA database (\texttt{edge\_pydb}) has been created to provide a homogeneous dataset derived from optical and millimeter observations \citep[][]{Wong2024}. 
Therefore, CALIFA and EDGE data are convolved to a common spatial resolution (FWHM$\sim$7 arcsec). Then the PIPE3D pipeline \citep{pipe3D1,pipe3D2} is applied to the convolved optical datacubes to produce two-dimensional maps of stellar and ionized gas properties at the same resolution as the CO surface brightness maps.
This database includes various estimates of CO moments from CARMA observations, including smoothed and masked CO datacubes. The \CO(1-0) cubes have a typical $\mathrm{3\sigma}$ sensitivity limit of $\mathrm{3 ~K ~km ~s^{-1}}$, which corresponds to $\mathrm{\Sigma_{mol}\sim13~ M_{\odot} ~pc^{-2}}$ \citep[][]{Wong2024}. 

We use the Square Grid Downsampling version of the EDGE-CALIFA database, where the data are sampled on a grid with 3~arcsec spacing in right ascension and declination, reducing the pixel count per galaxy from $128^2$ to $43^2$. This still oversamples the FWHM$\sim7$~arcsec Gaussian beam, covering each beam area with more than six sampling pixels. The database contains image data and CO spectra for 125 galaxies, with the downsampled data stored in the \texttt{edge\_carma.2d\_smo7.hdf5} and \texttt{edge\_carma.cocube\_smo7.hdf5} files, which provide CO and optical properties at matched resolution.
Specifically, we use the maps of the star formation surface density ($\mathrm{\Sigma_{SFR}}$), which trace the recent star formation activity, the stellar surface density ($\mathrm{\Sigma_{*}}$), representing the accumulated stellar mass over longer timescales, and the molecular gas surface density ($\mathrm{\Sigma_{mol}}$), which trace the current fuel available for star formation, for each galaxy explored in this study.
For a detailed explanation of how these observables are derived or the databases from which they were obtained, please refer to \citet{Wong2024}.

\subsection{ Main features of the sample}
\label{subsec:Main_characteristics}

For this analysis, we selected galaxies from the EDGE-CALIFA sample that have spaxels with $\mathrm{\Sigma_{mol}\geq13~M_{\odot}~pc^{-2}}$ to ensure reliable measurements, as detections below this threshold are statistically uncertain. Additionally, these spaxels are classified as star-forming regions, i.e., regions with an \Ha\, equivalent width (EW(\Ha)) greater than 6~\AA\ \citep{SanchezAnnRev2020} and that fall below the Kewley demarcation line \citep{Kewley2001} in the $\mathrm{[OIII]/\Hb}$ and $\mathrm{[NII]/\Ha}$ emission line ratio diagnostics \citep[BPT diagram;][]{Baldwin1981}. 

The selection criteria for the different merger stages are based on the scheme introduced by \citet{Veilleux2002}. 
This method assesses galaxy interaction/merging stages based on morphological classification using numerical simulations of spiral-spiral merger \citep{Surace1998}. 
Notably, this scheme describes the merger stages, which should be regarded as an indicator rather than a strict and definitive chronological sequence. Due to its inherent limitations, which derive from the nature of galactic interactions and observational capabilities.
The most important limitation is the uncertainty of classification. For instance,  morphological distortions during key merger stages can lead to an incorrect visual classification, such as post-merger stages being visually classified as pre-mergers \citep[][]{Veilleux2002}. In addition, merger remnants exhibit diverse morphologies (from spherical to disk-shaped), which are shaped by factors such as mass ratio, gas fraction, and orbital configurations \citep[e.g.,][]{Hopkins2009, Moreno_2019, Moreno2021}. Also, the identification of tidal features is limited by the brightness of the images, and such features may arise from both major and minor mergers \citep[e.g.,][]{Ebrova2013, Bilek2013, Bilek2022}.
However, this classification has been successfully applied to a subsample of CALIFA galaxies \citep[e.g.,][]{Barrera2015,Garay-Solis2023,Garay-Solis2024}. 
This allowed for homogeneous studies by applying the same observation and analysis methods to all the galaxies explored because it provides clear visual criteria for identifying each stage of merger.
Following these studies, we classify galaxies based on their  features observed in the $r$-band Sloan Digital Sky Survey \citep[SDSS,][]{SDSS_2000} at four different merger stages:

\begin{enumerate}
    \item Pre-merger: in this stage, the galaxy has a companion, but both galaxies show undisturbed morphologies, either spiral or elliptical. The main criteria used to classify this stage are: (1) the projected separation between each galaxy is less than 160 kpc \citep[][]{Patton2013,Barrera2015}, (2) systemic velocity difference under $\mathrm{300\, km\, s^{-1}}$ \citep[][]{Ellison2008,Ellison2013,Barrera2015}, and (3) difference in magnitudes in the $r$-band of less than 2 magnitudes \citep[][]{Barrera2015}.

    \item Merging: the galaxies exhibit features like tidal tails, bridges, and plumes while maintaining distinct nuclei \citep[][]{Veilleux2002,Barrera2015}.
      
    \item Post-merger:  galaxies with merged nuclei and prominent tidal features. Emissions from the core may be diffuse and obscured by dust lanes \citep[][]{Veilleux2002,Barrera2015}.

    \item Merger remnant: in this stage, the galaxy presents features such as weak tidal tails, shell shapes,  ripples, and fused cores. \citep[][]{Veilleux2002,Barrera2015}.
    
\end{enumerate}
  
Our sample includes 33 (1164 spaxels) in the isolated stage and 20 (659), 5 (55), 5 (158), and 4 (79) galaxies (spaxels) in the pre-merger, merging, post-merger, and merger remnant stages, respectively.
This sample consists mostly of spiral galaxies (Sa to Sd), covering a range of total stellar masses of $10^{9.5}-10^{11.4}~\Msun$,  global star formation rates of $10^{-1.0}-10^{1.0}~\Msun~$yr$^{-1}$, and total molecular masses of $10^{7.5}-10^{10.5}~\Msun$ \citep{Wong2024}.

\subsection{Characterizing the physical properties}
\label{subsec:characterizing_SFG}

\begin{figure*}
\centering
    \includegraphics[width=1.0\linewidth]{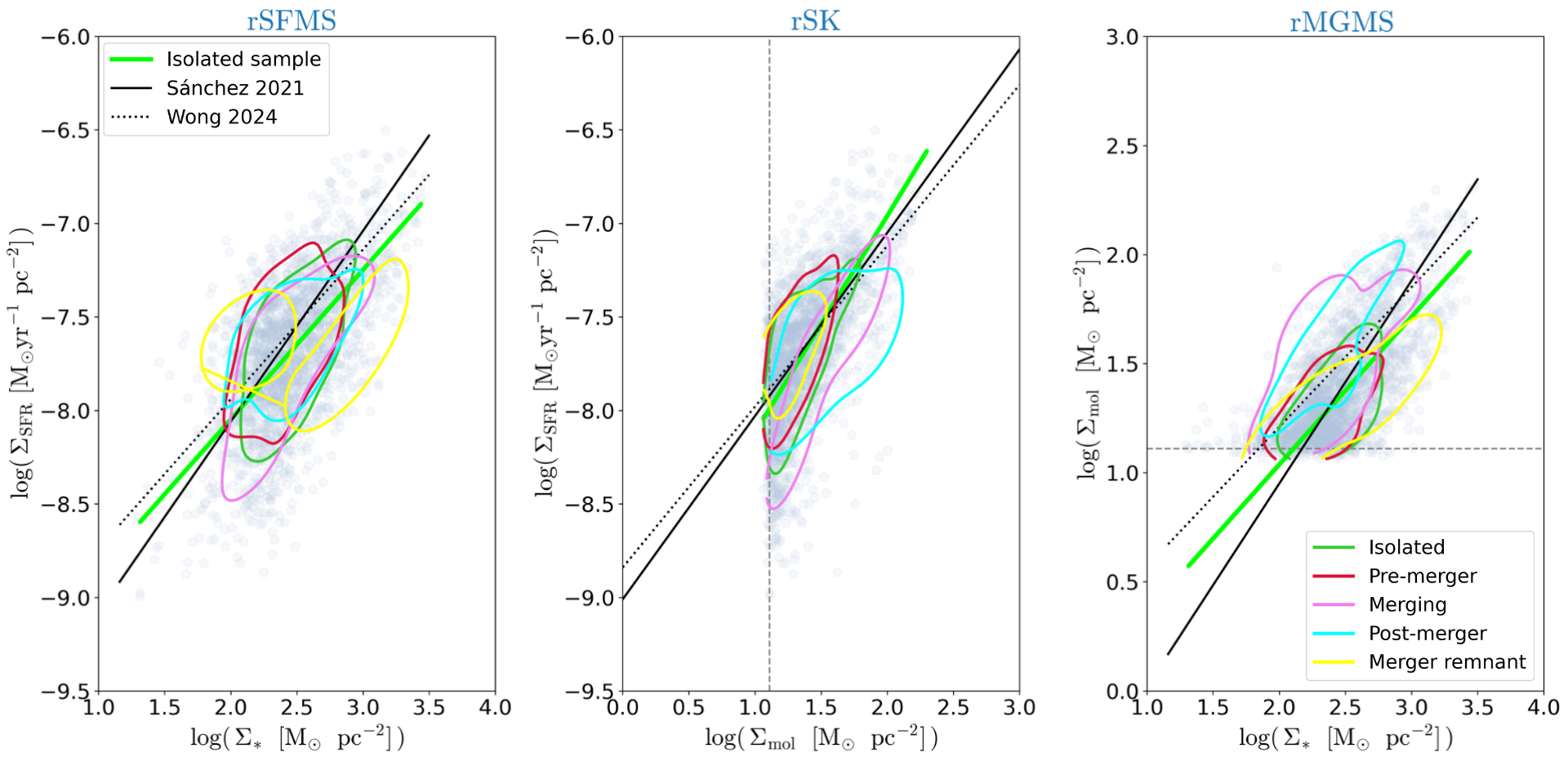}
    \caption{Distribution of the entire explored sample (2,115 star-forming regions) in the $\Ssfr - \Sstar$ (left panel), $\Ssfr - \Smol$ (middle panel), and $\Smol - \Sstar$ (right panel) diagrams. The green lines represent the best fit for the data from our isolated sample. The solid and dotted black lines show the fits derived by \citet{Sanchez2021} and \citet{Wong2024}, respectively. For comparison, we show $\sim68\%$ of the data from our isolated sample (green contour), along with the four different merger stages, from pre-merger to the merger remnant (red, pink, cyan, and yellow contours, respectively). We restricted our sample to spaxels within the $3\sigma$ detection limits of the CO observations. These limits are indicated by dashed lines in the middle and right panels.}  
    \label{fig:Scaling_relation}
\end{figure*}

To further explore the impact of interactions and mergers on the internal properties of galaxies, we will examine the behaviour of star formation-related properties at different radii in galaxies at different stages of the merger process. For this purpose, we analyse the following three spatially resolved scaling relations associated with the variation of star formation in galaxies. 
(1) The resolved Star-Formation Main Sequence (rSFMS): different studies have shown that star-forming regions exhibit a tight relation between the $\mathrm{\Sigma_{SFR}}$  and the $\mathrm{\Sigma_{*}}$ at kiloparsec scales, using mostly IFS data for galaxies in the local Universe \citep[e.g.,][]{Brinchmann2004,Cano2016,gonzalez2016,Ellison2018,Pan2018,Cano2019,Lin2019,SanchezAnnRev2020,Ellison2021b,Sanchez2021,Wong2024,Villanueva2024}.
(2) The resolved Schmidt–Kennicutt law (rSK law): is the relation that exists between the $\mathrm{\Sigma_{SFR}}$  and the $\mathrm{\Sigma_{mol}}$ with typically a slope near to 1 and a dispersion around $\sim$0.2 dex in $\mathrm{\Sigma_{SFR}}$ \citep[e.g.,][]{Wong2002,Kennicutt2007,Leroy2013,Bolatto2017,SanchezAnnRev2020,Ellison2021b,Sanchez2021,Kaneko2022,Wong2024,Villanueva2024}.  
(3) The resolved Molecular Gas Main Sequence (rMGMS): the relation between $\mathrm{\Sigma_{mol}}$ and $\mathrm{\Sigma_{*}}$ is characterized by a power-law with a slope of $\sim~1$ and a scatter $\sim~0.2$ dex \citep[e.g.,][]{Lin2019,Barrera2020,Ellison2021b,Sanchez2021,Wong2024}.

In Fig.~\ref{fig:Scaling_relation}, we present the aforementioned scaling relations for the 2,115 star-forming regions for our sample of 67 galaxies. From left to right: rSFMS, rSK, and rMGMS. 
In each panel, the total star-forming regions are shown in grey. To compare the distribution between different merger stages in these scaling relations, the colour-coded contours represent $\sim68\%$ of the data from isolated to merger remnant stages. 
We follow a similar procedure to that performed in previous studies to characterize these relations \citep[e.g.,][]{Sanchez2017,Barrera2017,Sanchez2021}. Specifically, we characterize these scaling relations by performing an Orthogonal Distance Regression (ODR) linear fit in the logarithmic scale for each one. To achieve this, we calculate the median values within a set of consecutive bins of 0.15~dex for the parameter of interest in each panel for our isolated sample. These binned medians are then fitted to the equation: $y = \beta + \alpha\ x$.
The green lines represent the best-fitting linear relation between each pair of parameters of the isolated regions. 
For the rSFMS: $\beta = -9.65\pm0.11$ and $\alpha = 0.80\pm0.05$; 
for the rSK: $\beta = -9.28\pm0.08$ and $\alpha = 1.16\pm0.06$; 
and for the rMGMS: $\beta = -0.32\pm0.15$ and $\alpha = 0.68\pm0.06$. 
These fits are in excellent agreement with those reported by \citet[][]{Sanchez2021} and \citet[][]{Wong2024},  represented by the solid and dotted black lines, respectively.
Specifically, \citet[][]{Sanchez2021} reported slopes of $1.02\pm0.16$ (rSFMS), $0.98\pm0.14$ (rSK), and $0.93\pm0.11$ (rMGMS); while \citet[][]{Wong2024} found $0.80$ (rSFMS), $0.86$ (rSK), and $0.64$ (rMGMS).
Our results fall well within the uncertainties of these previous works.

In the rSFMS, we observe that at kiloparsec-scales the pre-merger, merger, and post-merger stages follow a similar trend to that of isolated galaxies. In the merger remnant stage, unlike the isolated sample, we find a tendency for the regions to cluster above and below the main sequence, with a breaking point at $\mathrm{log~ \Sigma_{\star}\sim2.4}$ dex. It should be noted that, at this stage, the small number of spaxels ($\sim 79$ regions) results in poor coverage of the parameter space, thus increasing the scatter with respect to the trends of main-sequence galaxies.

In the rSK, we find that the pre-merger and merger remnant stages follow a similar trend to the isolated sample, exhibiting a comparable dispersion. In the merging and post-merger stages,  we find that the regions tend to cluster below the best fit, with some regions farther from the best fit.

In the rMGMS relation, we find that resolved regions in pre-merger galaxies show a similar trend to the isolated sample. In the merging and post-merger stages, unlike the isolated sample, the regions show greater dispersion, clustering above and farther from the best fit, i.e., at a given $\mathrm{\Sigma_{\star}}$, these regions exhibit a higher $\mathrm{\Sigma_{mol}}$. In the merger remnant stage, we find that the regions show greater dispersion, clustering below the best fit.

To quantify the variations in the specific Star Formation Rate ($\mathrm{sSFR=\Sigma_{SFR}/\Sigma_{\star}}$), the Star Formation Efficiency ($\mathrm{SFE=\Sigma_{SFR}/\Sigma_{mol}}$), and the molecular gas fraction ($\mathrm{f_{mol}=\Sigma_{mol}/\Sigma_{\star}}$), we derive the residuals of each scaling relation. These residuals trace the increase or decrease of these properties relative to the expected trend \citep{Garay-Solis2023}. Specifically, the residuals are defined as $\Delta {\mathrm{SFMS}} = \log(\Sigma_{\mathrm{SFR}}) - (\alpha  \log(\Sigma_{\star}) + \beta)$, $\Delta {\mathrm{SK}} = \log(\Sigma_{\mathrm{SFR}}) - (\alpha  \log(\Sigma_{\mathrm{mol}}) + \beta)$, and $\Delta {\mathrm{MGMS}} = \log(\Sigma_{\mathrm{mol}}) - (\alpha \log(\Sigma_{\star}) + \beta)$, for rSFMS, rSK, and rMGMS, respectively. 
To account for individual measurement uncertainties, we performed 1000 Monte Carlo iterations. All results presented in this study are the averages of these iterations.

\subsection{Parameters derived from random data}
\label{subsec:mock_data}

To assess whether the correlations between the residuals \Dsfms, \Dsk, and \Dmgms\ have a statistical origin, we generate a random dataset following a procedure similar to that described in previous studies \citep[e.g.,][]{Barrera2021, Sanchez2021, Garay-Solis2023, Garay-Solis2024}. This dataset (hereafter referred to as mock data) consists of $\sim200000$ values simulating regions characterized by a \Sstar, a \Ssfr, and a \Smol\ at different radii for 1000 mock galaxies.

In order to derive these mock data, we first generate a random distribution of 1000 total stellar mass values (\Mstar). This distribution is designed to resemble that of the observed in our isolated sample. Then, we generate the radial distribution of mock \Sstar, accounting for all morphologies and grouping galaxies based on their \Mstar. To achieve this, we use the fits and characterizations of the radial distribution of the \Sstar\, derived by \citet[][]{Barrera2023}.
We use the best fit of the rSFMS to derive the mock \Ssfr\ based on the mock \Sstar.
Similarly,  we estimate the mock \Smol\ using the best fit of the rMGMS. Using these mock \Smol, we also derive the mock \Ssfr\ based on the best fit of the rSK. The final mock \Ssfr\ is derived as the average of the estimates from the rSFMS and the rSK.
To account for the systematic uncertainties of the observations, we introduce perturbations to the estimated values of mock properties based on their observational errors, i.e., $\sim0.03$~dex, $\sim0.18$~dex, $\sim0.28$~dex, and $\sim0.15$~dex for \Sstar, \Smol, \Ssfr, and radius, respectively.

We also perform the mock scaling relations and we calculate the mock residuals following the same procedure used for the observational sample, using the best fit of their respective scaling relations (see Section~\ref{subsec:characterizing_SFG}).
In the next section, we compare the radial distributions of the residuals from mock data with observations to assess the impact of interactions and mergers on internal properties at different radii, determining whether the behaviour found is due to a physical or statistical origin.

\section{Results}
\label{sec:Results}
\subsection{SFE and $\mathbf{f_{mol}}$ variations in the Merger Process}
\label{sec:DeltaSK_DeltaMGMS}

\begin{figure*}
\centering
    \includegraphics[width=0.9\linewidth]{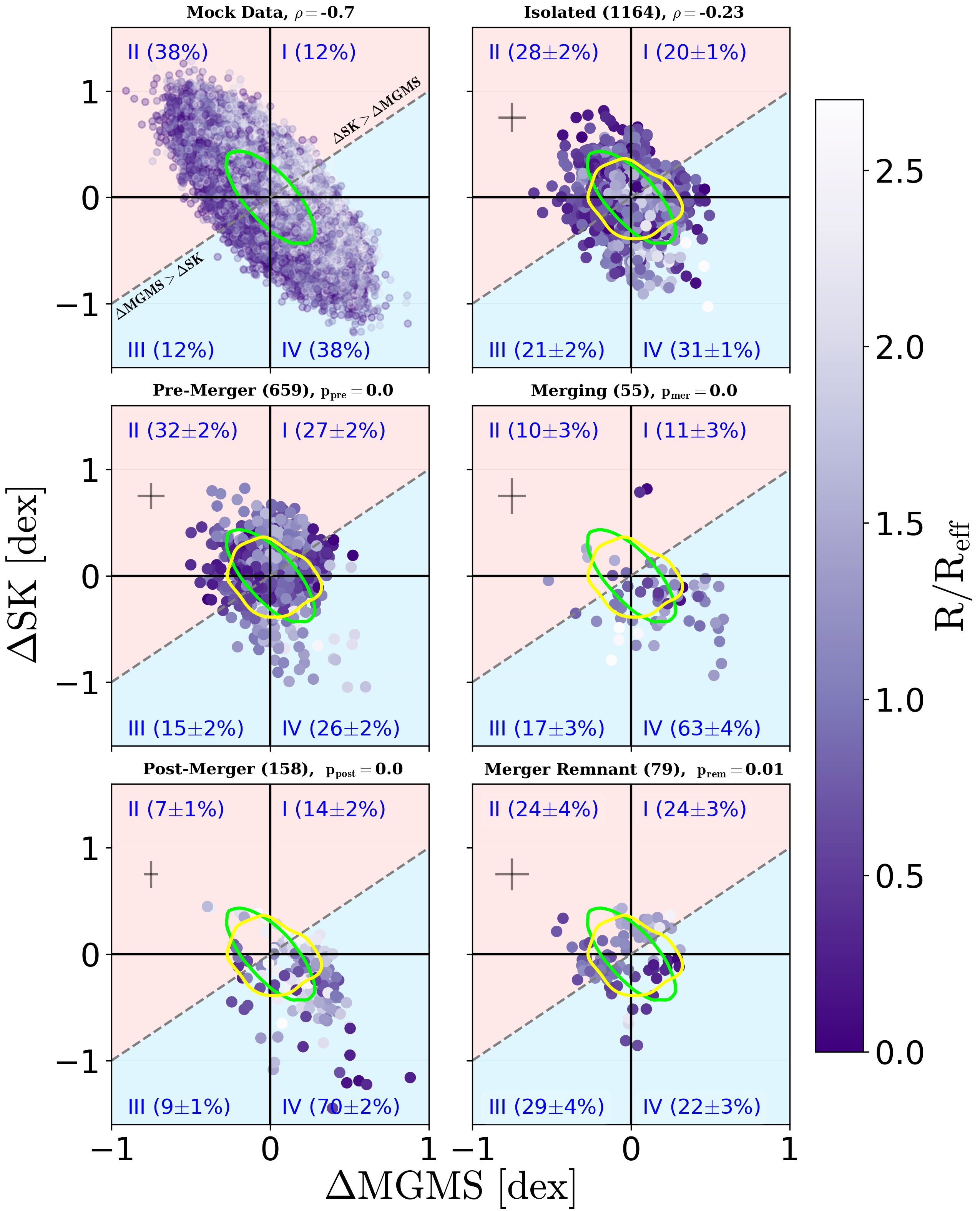}
    \caption{\Dsk\ against \Dmgms\ colour-coded by $R/\Reff$. Each panel shows the number of star-forming regions explored at different stages of the merger process. $\mathbf{\rho}$ indicates the Spearman correlation coefficient for the mock data and isolated sample (top panels). $\mathbf{p}$ represents the $p$-value from the two-dimensional Kolmogorov-Smirnov (KS-test) between the different merger stages and the isolated sample. The green and yellow contours encircle $\sim68\%$ of the mock data and the isolated sample, respectively. The quadrants I, II, III, and IV indicate in parentheses the fraction of regions within each quadrant.  The error bars in quadrant II indicate the standard uncertainty associated with the residuals in each panel. The light-red area highlights regions where \Dsk\ is greater than their \Dmgms, and the sky-blue area highlights the opposite ($\Dmgms>\Dsk$). }
    \label{fig:Stages_radio}
\end{figure*}

In this section, we explore the relative increase or decrease of the SFE and the \fg\ across the merger sequence. As we mentioned in Section~\ref{subsec:characterizing_SFG}, these properties are traced by the residuals \Dsk\ and \Dmgms, respectively.
Therefore,  in Fig.~\ref{fig:Stages_radio} we present the distribution of the residuals \Dsk\ against \Dmgms\ colour-coded by galactocentric radii ($R/\Reff$).
We plot this $\Dsk-\Dmgms$ diagram for the different merger stages, including mock data and isolated sample. Following \citet[][]{Moreno2021} and \citet[][]{Garay-Solis2023}, each panel is divided into four quadrants (labelled I through IV). The percentages in parentheses indicate the fraction of each sample located in each quadrant. The total number of regions in each stage is indicated at the top of each panel.

We find that the mock and isolated sample distributions cover similar areas, as indicated by the contours enclosing $\sim68\%$ of the simulated data (green) and the isolated samples (yellow). However, although both samples exhibit an anticorrelation, this is weaker in the isolated sample. This can be seen by the Spearman correlation coefficients, with $\rho \sim-0.70$ for the mock data and $\rho \sim-0.23$ for the isolated sample.  In agreement with previous studies \citep[e.g.,][]{Sanchez2021,Garay-Solis2023}, this analysis suggests that the correlations between residuals are driven by data stochasticity rather than by a physical mechanism. To quantify the similarity between the mock data and the isolated samples, we derive the two-dimensional Kolmogorov-Smirnov (KS-test), obtaining a $p-$value ($p = 0$).  This value suggests that both samples are significantly different. However, $\sim50\%$ of each sample presents an increase in their \fg\ (regions located in quadrants I and IV) and SFE (regions located in quadrants I and II). This provides evidence that the two samples are similar, which is consistent with the anticorrelation of statistical origin. On the other hand, the mock data and the isolated sample distributions do not show a clear trend as a function of galactocentric radius. In Section~\ref{subsec:profiles_Dsk} and Section~\ref{subsec:profiles_Dmgms}, we explore in more detail the impact of mergers on these properties at different radii.

In order to evaluate whether the distributions of each merger stage and the isolated sample are statistically similar, we use the $p$-value from the KS-test. We find that the four samples (pre-merger to merger remnant stage) exhibit a $p$-value very close to zero. This suggests that the differences observed between each merger stage and the isolated sample are unlikely to be due to random chance. It indicates that the interaction or merging process has a significant and measurable impact on the properties analysed. Unlike the isolated sample, we find that, in the merging and post-merger stages, a significant fraction of the regions is located in quadrant IV ($\sim63\%$ and $\sim70\%$, respectively, against $\sim31\%$), i.e., these regions exhibit an enhancement in their \fg\ ($\Dmgms > 0$) but a diminishment in SFE ($\Dsk < 0$).  It is important to note that in the post-merger sample, the outliers of quadrant IV are located closer to the galaxy centre ($R<0.5~\Reff$). On the contrary, in the merging sample, the outliers are located farther to the galaxy centre ($R>1.5~\Reff$).
For the pre-merger and merger remnant samples, we find similar fractions to those of the isolated sample in regions located in quadrant IV ($\sim26\%$ and $\sim22\%$, respectively). Also, in quadrant IV, most regions are located within $R<0.5~\Reff$.
These results indicate that the interaction/merger process significantly impacts the SFE and \fg, particularly in the merging and post-merger stages, leading to a non-uniform redistribution of molecular gas across different radii within galaxies.
In general, our results are consistent with previous studies that analysed data integrated at $\sim 1~\Reff$ \citep[][]{Garay-Solis2023}.

\subsection{The primary driver of the enhanced and suppressed star formation}
\label{sec:Main_Drivers}

\begin{figure}
\centering
    \includegraphics[width=1\linewidth]{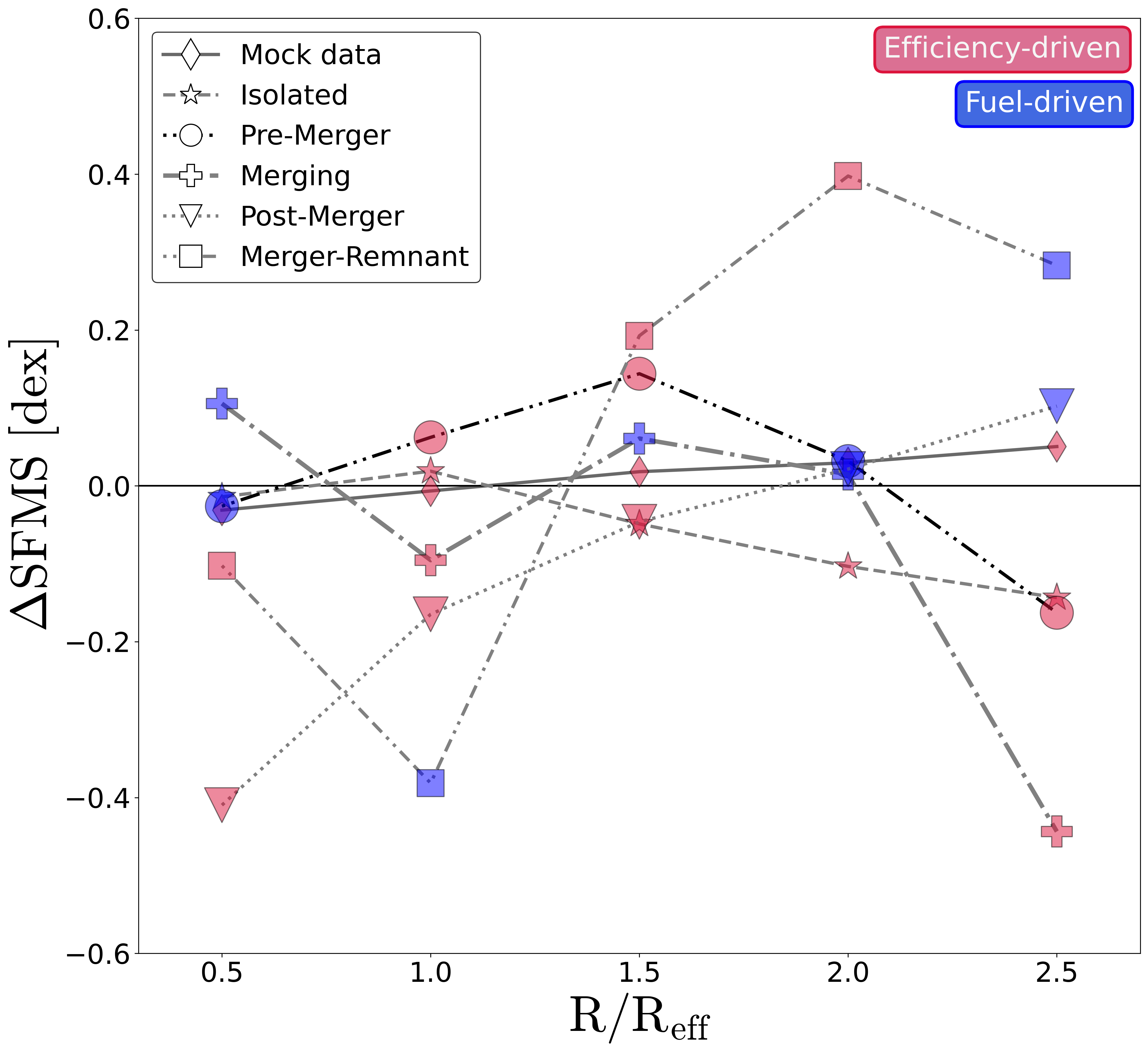}
    \caption{
    Distribution of the \Dsfms\ median in bins of $0.5~\Reff$ for each merger stage. Each symbol is colour-coded to indicate whether the increase ($\Dsfms>0$) or decrease ($\Dsfms<0$) in sSFR is primarily driven by the molecular gas (fuel) or by the efficiency, blue and red colours, respectively.
    }
    \label{fig:Stages_driver_radio}
\end{figure}

In this section, we analyse how the merging process affects the primary driver of excess (deﬁcit) in star formation at different radii. In other words, whether the enhanced (diminished) sSFR is driven by efficiency or molecular gas (fuel). To achieve this, we study the main driver of the \Dsfms\ median within radial bins. Following \citet[][]{Moreno2021} and \citet[][]{Garay-Solis2023}, we divide the regions into two groups: (1) regions with excess \fg\ relative to SFE ($\Dmgms > \Dsk$), located in the sky-blue area in Fig.~\ref{fig:Stages_radio}; and (2) regions with excess SFE relative to \fg\ ($\Dsk > \Dmgms$), located in the light-red area in Fig.~\ref{fig:Stages_radio}. The main driver of sSFR can be defined by comparing the fractions of regions in these groups within a given \Dsfms\ bin. If (1) exceeds (2), the sSFR enhancement (diminishment) is fuel (efficiency)-driven. On the contrary, if (2) is larger than (1), the enhanced (suppressed) sSFR is efficiency (fuel)-driven.

In Fig.~\ref{fig:Stages_driver_radio}, we present the distribution of \Dsfms\ median for the regions contained in bins of $0.5~\Reff$.
The symbols show the mock data sample, the isolated sample, and the four merger stages, respectively. 
Blue colour indicates that the increase or decrease in sSFR is fuel-driven, and the red colour indicates that it is efficiency-driven.

In the mock data (diamond), we find that both the enhancement ($\Dsfms>0$) and diminishment ($\Dsfms<0$) of sSFR are efficiency-driven across all the radial bins. 
In the isolated sample (stars) at $R<0.5~\Reff$, unlike mock data, the main driver of the diminished sSFR is fuel.
At $0.5<R<1.0~\Reff$, the main driver of the increase in sSFR is efficiency. 
Also, we find that the main driver of the decrease in star formation is the efficiency for $R>1.0~\Reff$.
In this sample, we find that the outer radii regions exhibit a more decreased sSFR than the inner ones, driven by efficiency, suggesting that a reduction in the ability to convert gas into stars is primarily responsible. 

In the pre-merger sample (circles), we find that, similar to the isolated sample, the diminished sSFR at $R<0.5~\Reff$ is fuel-driven. 
However, at radii of 0.5 to $1.5~\Reff$, the sSFR is more than the isolated sample, and its main driver is efficiency-driven. 
At $1.5 < R < 2.0~\Reff$, the molecular gas is the main driver of sSFR enhancement.
The suppressed sSFR observed at $R > 2.0~\Reff$ is efficiency-driven, similar to the isolated sample.
These results suggest that specific regions of galaxies are significantly and differently affected by the interactions.

In the merging sample (plus signs), we find behaviours different from those of the isolated sample.
This stage exhibits the largest increase in sSFR driven by fuel within $R<0.5~\Reff$. 
However, we find that the diminished sSFR is efficiency-driven at $0.5<R<1.0~\Reff$.
For $1.0<R<2.0~\Reff$, we find that the main driver of sSFR enhancement is fuel, whereas the diminished sSFR is mainly driven by efficiency at $R > 2.0~\Reff$,
In general, the enhanced star formation is driven by the amount of molecular gas. However, the efficiency determines suppressed star formation. 
In the post-merger sample (triangles), we find that the diminished sSFR is driven by efficiency for $R<1.5~\Reff$. While for $R>1.5~\Reff$, we find that enhanced sSFR is fuel-driven.
In the merger remnant sample (squares), we find that, unlike the isolated sample, at $R<0.5~\Reff$, the diminished sSFR is efficiency-driven. 
In contrast, at $0.5<R<1.0~\Reff$ the diminished sSFR is mainly fuel-driven.
At $1.0<R<2.0~\Reff$, the sSFR enhancement is efficiency-driven, but it is efficiency-driven for $R>2.0~\Reff$. 

Our results suggest that the main driver of star formation activity changes in a complex manner depending on both the merger stage and the galactocentric radius. 
Therefore, in Appendix~\ref{Appendix_fractions}, we present the primary driver of both enhanced and suppressed star formation for different bins of \Dsfms\ and radius, finding that regardless of the amount of molecular gas available, efficiency appears to play a crucial role in both the enhancement and suppression of star formation.

\subsection{The radial  distribution of $\mathbf{\Delta SFMS}$}
\label{subsec:profiles_Dsfms}

\begin{figure}
\centering
    \includegraphics[width=1\linewidth]{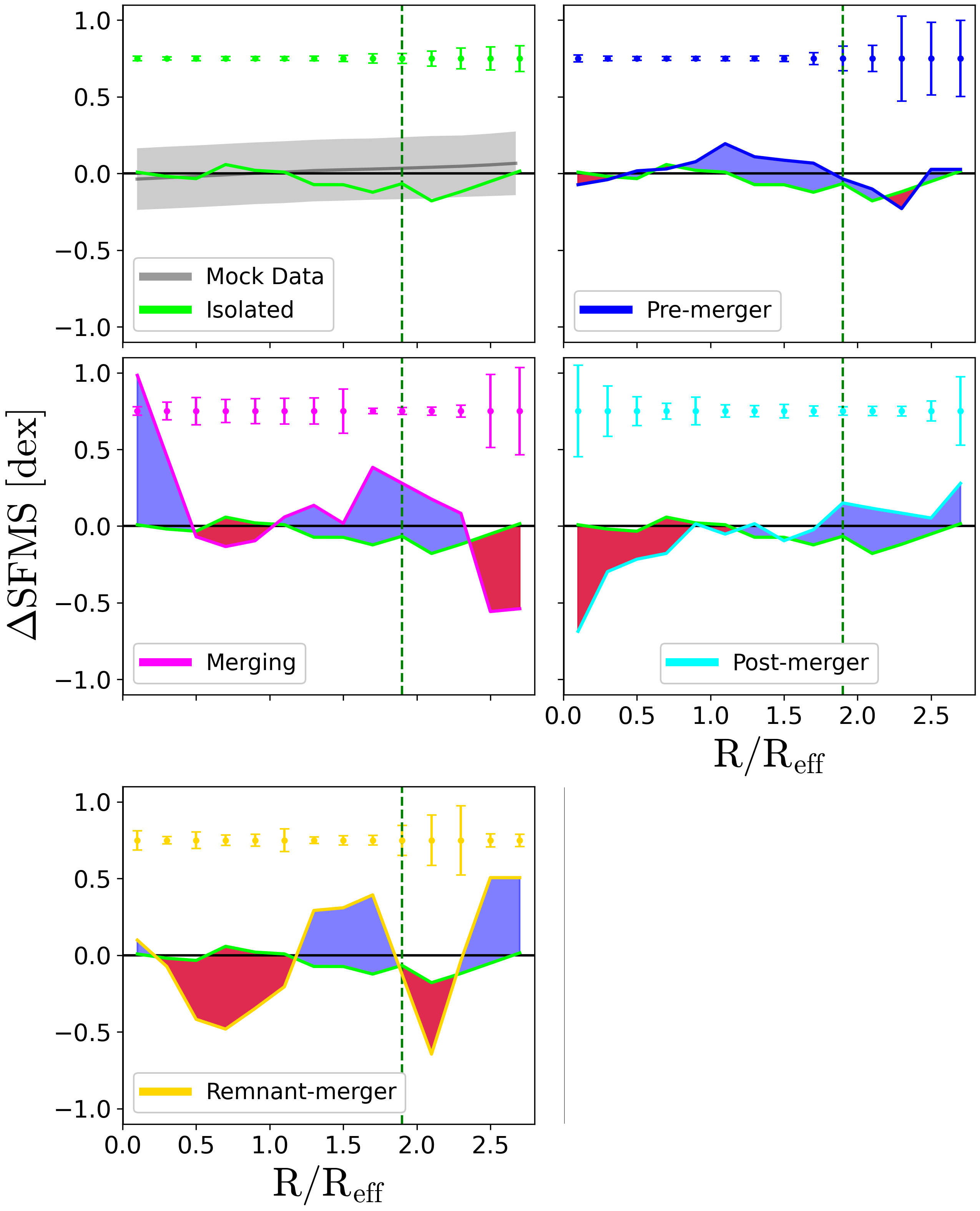}
    \caption{Radial distribution of the Star Formation Main Sequence residuals (\Dsfms) for the different merger stages. In the top-left panel, we show the profile of our isolated sample and the mock sample (green and grey, respectively). In the other panels, we show the profiles of the four different merger stages in comparison with our isolated sample (green line). Radial profiles are shown as a function of the effective radius, with the solid line indicating the median in $0.2~\Reff$ bins, and the top bars indicating the error on each median. The vertical dotted line indicates the reliability limit of the profile for our isolated sample, as at radii smaller than $1.9~\Reff$, each bin includes data from at least 5 galaxies. The purple region indicates the radii where the merger stage shows an increase in the sSFR, traced by \Dsfms, compared to the control (isolated) sample. The red region indicates the opposite, i.e., the radii where the sSFR shows a decrease in comparison to the control sample.}
    \label{fig:profiles_SFMS}
\end{figure}

In order to explore how interactions impact sSFR variations at different radii, in Fig.~\ref{fig:profiles_SFMS}, we present the median profiles of \Dsfms\ in $0.2~\Reff$ bins. 
In the top-left panel, we plot the radial distribution of \Dsfms\ of the mock sample (grey line) and the isolated sample (green line).
Since each galaxy covers different radii, the median per bin is calculated from data of at least five galaxies, with this condition being met up to $1.9~\Reff$ (vertical dotted line).
Therefore, we consider this profile reliable up to this radius. 

As expected, we find a flat radial profile for the mock data since we build this sample by adding random values to the best fit.
This suggests that the trends observed in the radial profiles are likely due to underlying physical processes and not just to the stochasticity of the data.
The isolated sample (green line) exhibits a behaviour similar to that of the mock sample. 
In other words, \Dsfms\ is close to zero in most radii, without significant variations. This suggests that, on average, isolated galaxies do not exhibit a systematic increase or decrease in the specific star formation rate (sSFR) as a function of galactocentric distance. Furthermore, the data dispersion, represented by the error bars, is relatively small, indicating that this trend is consistent throughout the sample.
In the other panels, we compare the radial distribution of the isolated galaxies with each of the different merger stages.
In order to visualize the radii where an increase relative to the isolated sample is exhibited, we highlight these regions in purple. In contrast, the regions that exhibit a decrease relative to the isolated sample are coloured red. 
In the pre-merger sample, at $R<0.9~\Reff$, we find that the \Dsfms\ are close to the isolated sample profile, showing no significant differences. However, at larger radii ($0.9<R<1.9~\Reff$), we find a slight tendency for \Dsfms\ to increase ($0.19$~dex) with respect to the profile of the isolated sample. This suggests that the pre-merger galaxies do not show a significant impact on their star formation in the central region, but these galaxies exhibit a mild enhancement in star formation in the intermediate and outer regions.
As mentioned above, we cannot reliably compare radii greater than $1.9~\Reff$, although the behaviour is very similar to that of the isolated sample.
In the merging sample, at $R<0.5~\Reff$, we find that the \Dsfms\ are significantly higher ($\sim1.00$~dex) than observed in isolated galaxies,  indicating enhanced star formation.
However, at $0.5<R<1.0~\Reff$, we find a slight decrease ($\sim-0.19$~dex); and for $1.0<R<1.9~\Reff$ the \Dsfms\ begins to increase to $\sim0.50$~dex. This suggests that mergers enhance star formation primarily in the innermost and outermost regions of galaxies.
In contrast, in the post-merger sample, we find that the \Dsfms\ decrease from $\sim-0.70$~dex to $\sim0.00$~dex for $R<1.2~\Reff$. 
For $1.2<R<1.9~\Reff$, the \Dsfms\ show a slight increase ($\sim0.22$~dex) compared to the isolated profile.
In the merger remnant sample, we find that in the central region ($R<0.2~\Reff$) the variations of \Dsfms\ are very close to the isolated ones, but at $0.2<R<1.2~\Reff$ the \Dsfms\ decrease to $\sim-0.54$~dex and at $1.2<R<1.9~\Reff$ the \Dsfms\ increase to $\sim0.52$~dex.
These results suggest that the impact of the mergers on star formation is not uniform across different galactocentric distances.

\subsection{The radial  distribution of $\mathbf{\Delta SK}$}
\label{subsec:profiles_Dsk}

\begin{figure}
\centering
    \includegraphics[width=1\linewidth]{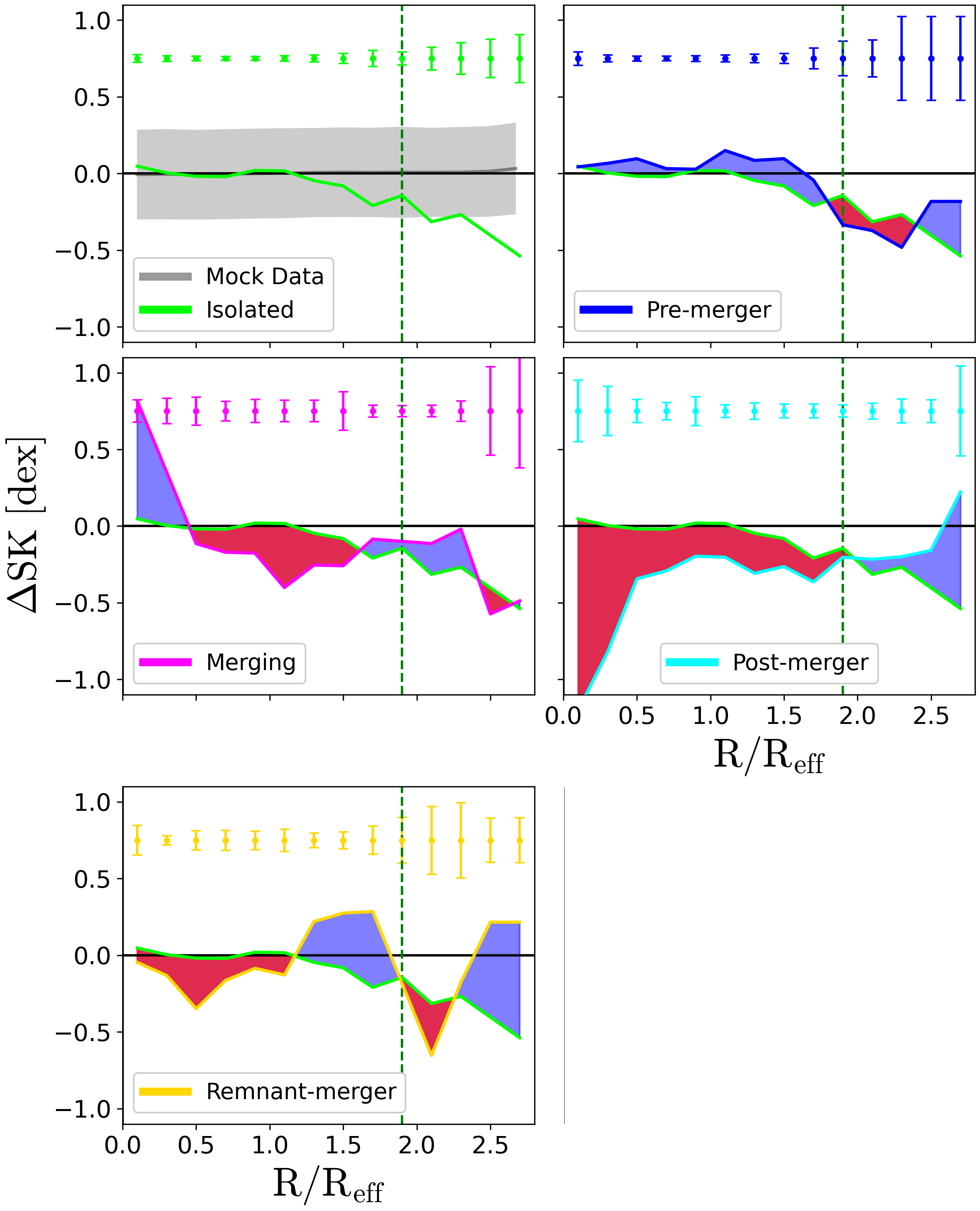}
    \caption{Following the scheme adopted in Fig.~\ref{fig:profiles_SFMS}, we show the radial distribution of the Schmidt-Kennicutt law residuals (\Dsk) for the different merger stages compared to the isolated sample.
    }
    \label{fig:profiles_SK}
\end{figure}

In Fig.~\ref{fig:profiles_SK}, we show the radial distribution of SFE (traced by \Dsk) for the different merger stages, following the same scheme adopted in Fig.~\ref{fig:profiles_SFMS}. For the isolated sample, we find that the profile of \Dsk\ exhibits a similar trend to the profile of \Dsfms\ for $R<1.9~\Reff$, with residuals close to zero, i.e., the efficiency of converting gas into stars does not show significant deviations from the expected trend. 
In the pre-merger sample, we find that the \Dsk\ are greater than the isolated profile ($\sim0.19$~dex), for $R<1.9~\Reff$. It indicates that galaxy interactions increase the SFE at different radii. 
However, we find that in galaxies in the merging stage, the \Dsk\ are increased in the inner ($R<0.5~\Reff$) and outer ($R>1.5~\Reff$) regions in comparison with the isolated profile, $\sim0.82$~dex and $\sim0.13$~dex, respectively. In the intermediate region ($0.5<R<1.5~\Reff$), the \Dsk\ are lower than those find in isolated galaxies ($\sim-0.40$~dex).
In the post-merger sample, at $R<1.9~\Reff$, we find a \Dsk\ profile more negative than the isolated profile, from $\sim-1.20$~dex to $\sim0.0$~dex. As we can see, the behaviour of the SFE changes dramatically from one stage to another.
In the merger remnant stage, we find a decrease in \Dsk\ ($\sim-0.34$~dex) for $R<1.2~\Reff$, while for $1.2<R<1.9~\Reff$, \Dsk\ increases $\sim0.50$~dex compared to the isolated profile.

These results suggest that the impact of galaxy interactions on SFE varies significantly across different merger stages, and it can change suddenly between different radii. This behaviour highlights the complex role of galaxy interactions/mergers in star formation.

\subsection{The radial  distribution of $\mathbf{\Delta MGMS}$}
\label{subsec:profiles_Dmgms}

\begin{figure}
\centering
    \includegraphics[width=1\linewidth]{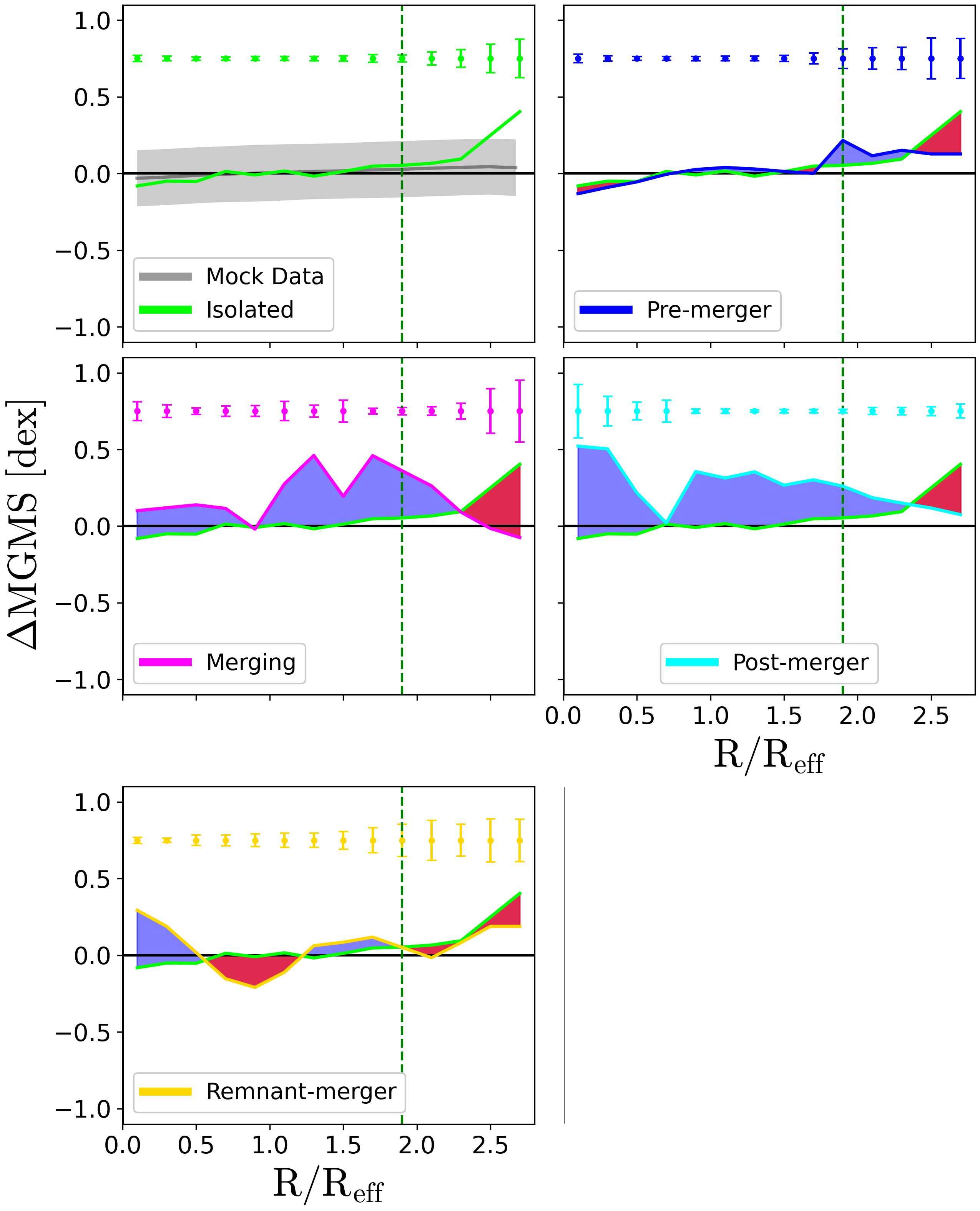}
    \caption{Radial distribution of the Molecular Gas Main Sequence residuals (\Dmgms) for the different merger stages. The panels follow the scheme adopted in Fig.~\ref{fig:profiles_SFMS}.}
    \label{fig:profiles_MGMS}
\end{figure}

In Fig.~\ref{fig:profiles_MGMS}, we present the radial profiles of \fg\ (traced by \Dmgms) across different merger stages, using the same scheme as in Fig.~\ref{fig:profiles_SFMS}. We find that isolated galaxies exhibit a \Dmgms\ profile similar to those of \Dsfms\ and \Dsk\ for $R<1.9~\Reff$, i.e., the molecular gas follows the expected distribution. 
In galaxies at the pre-merger stage, we find a behaviour of \Dmgms\ very similar to that of isolated galaxies. This suggests that galaxies at this stage do not exhibit significant variations in the molecular gas fraction across different radii. Considering that at $0.9 < R < 1.9~\Reff$, pre-merger galaxies exhibit an increase in \Dsfms, compared to the isolated sample, but do not show an increase in molecular gas, an increase in SFE drives this enhancement in star formation.
For merging galaxies at $R<1.9~\Reff$, we find a \Dmgms\ profile up to $\sim 0.47$~dex higher than the isolated profile. At $R<0.9~\Reff$ the \Dmgms\ are smaller ($\sim 0.19$~dex). However, despite this small increase in the \fg, we find the largest increase in \Dsfms\ ($\sim 1.00$~dex) for $R<0.5~\Reff$, which is driven by efficiency. This suggests that the inner regions of these galaxies are more efficient at converting molecular gas into stars. However, in the intermediate and outer regions, despite the accumulation of more molecular gas, the sSFR does not increase as much as in the inner regions, and this appears to be driven by efficiency.
For post-merger galaxies, we find a behaviour opposite to that of merging galaxies. In these galaxies, we also find a \Dmgms\ profile up to $\sim 0.58$~dex higher than the isolated profile, but the largest residuals ($\sim 0.58$~dex) occur at $R<0.5~\Reff$. 
As we mentioned in Section~\ref{subsec:profiles_Dsfms} and Section~\ref{subsec:profiles_Dsk}, the \Dsfms\ profile shows only decreases and the \Dsk\ profile exhibits negative values for $R<1.2~\Reff$. These results suggest that the galaxies in the post-merger stage, although molecular gas increases at different radii, this is not sufficient to enhance sSFR. Also, the primary driver of suppressed SFR is efficiency, consistent with previous studies \citep[][]{Garay-Solis2023}. 
In the merger remnant sample, we find increases in \Dmgms\ at both inner ($\sim0.37$~dex for $R<0.5~\Reff$) and outer ($\sim0.09$~dex for $1.2<R<1.9~\Reff$) regions, while intermediate regions ($0.5<R<1.2~\Reff$) exhibit decreases in \Dmgms\ of $\sim-0.20$~dex. 
As shown in Fig.~\ref{fig:profiles_SFMS} and Fig.~\ref{fig:profiles_SK}, at $R<1.2~\Reff$, there is a deficit in the sSFR and the SFE, respectively. This applies to both the inner region, where \fg\ increases, and the intermediate region, where \fg\ decreases. That is, regardless of whether there is an excess or a deficit of molecular gas, star formation does not increase.
In the outer regions, the opposite is observed; we find a $\sim0.52$~dex increase in sSFR, although there is a slight increase of $\sim0.09$~dex in \fg. 

Our results indicate that, as a consequence of the merger process, galaxies exhibit an increase in molecular gas during the merging, post-merger, and merger remnant stages compared to isolated galaxies. However, these increases do not show a clear or uniform trend with galactocentric distance and vary significantly across different stages, such as between the merging and post-merger stages.

\section{Discussion}
\label{sec:discusion}

In this study, we investigate how the merging process impacts the internal properties of galaxies from the EDGE-CALIFA survey at different radii. Specifically, we analyse the variations in SFE,  \fg, and sSFR, as well as the primary drivers of the latter.
Previous investigations have carried similar objectives. For instance, recently \citet{Thorp_2024} studied star formation rate profiles in 150 post-merger galaxies from the MaNGA \citep[][]{Manga} survey, focusing on galaxies with starbursts. They found that post-merger galaxies exhibit a significantly stronger central starburst compared to isolated galaxies with similarly enhanced global star formation ($R<\sim 0.5~\Reff$).
\citet{Thorp2019} explored SFR profiles in 36 post-merger galaxies from the MaNGA survey as well. They found that SFR enhancements in post-merger galaxies tend to be centrally concentrated but can also extend to larger radii, with variations in the outer regions, which may exhibit both enhancement and suppression of star formation.
In this work, we do not find significantly enhanced central star formation for our galaxies in the post-merger stage (see Section~\ref{subsec:profiles_Dsfms}). However, we find a star formation enhancement in the outer regions of galaxies at this stage. 
This last result is consistent with those reported by \citet{Thorp2019} for the outer radii.
However, the disagreement between our results and those of \citet{Thorp_2024} may be attributed to the fundamental differences in the scope and methodology of each work to achieve the different scientific objectives. On the one hand, our study utilizes a smaller sample of post-merger galaxies (5 vs. 150). On the other hand, unlike \citet{Thorp_2024}, our analysis includes galaxies both above and below the global SFMS, and we compare each merger stage with the rSFMS derived from our isolated sample.

\citet{Pan2019} investigated SFR in 205 interacting and merging star-forming galaxies from the MaNGA survey. Their findings indicate that interactions have little impact during the incoming stage. After the first pericentre passage, there are transitions from enhanced to suppressed sSFR as the radius increases. At later stages, sSFR is broadly enhanced, peaking centrally and extending up to $\sim6.7$ kpc. Later, the increased SF is also observed at apocenters and during coalescence. 
Also, \citet{Steffen2021} compare the radial profiles of sSFR in 169 star-forming galaxies in close pairs from the MaNGA survey. They found that galaxy interactions lead to enhanced central star formation ($R<1~\Reff$) and suppressed outer sSFR.
The findings of these works are in agreement with our results since we find a slight star formation enhancement for our galaxies in the pre-merger stage.  
At the merging galaxies, we find a change from enhanced to suppressed sSFR and vice versa as the radius increases.
We do not find significantly enhanced central star formation for our galaxies in the post-merger and merger remnant stages. However, we find a star formation enhancement in the outer regions of galaxies in our different merger stages. Our results indicate that the star formation enhancements due to the merger process extend beyond the central region.

Previous studies have investigated how mergers influence the drivers of star formation at kiloparsec scales. For instance, \citet{Thorp2022} analysed 31 pair/post-merger galaxies, and \citet{Ellison2020} explored 3 recently merged galaxies; both studies included galaxies from the ALMaQUEST \citep[ALMA MaNGA QUEnching and STar formation;][]{Lin2020} survey. They found that regardless of the interaction stage, different regions within a galaxy may exhibit different drivers of star formation. However, in the central region, efficiency appears to be the main driver in both enhanced and suppressed star formation. 
These results are consistent with our findings, as, in general, although we observe an increase in molecular gas at different radii in galaxies during the merger stages, both enhanced and suppressed star formation are mainly driven by efficiency. 
On the other hand, while \citet{Colombo2020} found that a lack of molecular gas is the primary driver for initiating quenching in a general galaxy population, our results highlight the overriding importance of SFE in merger events. We find that during the late stages of a merger process, where quenching is expected to begin \citep[e.g.,][]{Petersson_2023, Hopkins_2008}, a galaxy can experience a decrease in SFR that can be driven by an efficiency deficit, even when fuel is available.

Regarding studies using simulations, \citet[][]{Faria2025} used IllustrisTNG cosmological simulations to explore the sSFR and the hydrogen star formation efficiency ($\mathrm{SFE_{H}}$) in galactic interactions.
Our results agree with theirs in that interactions redistribute and increase the amount of gas available for star formation, and that the efficiency of this process is affected.
However, \citet[][]{Faria2025} emphasize the temporal delay of the sSFR peak after pericenter passage, while we highlight the non-uniformity of these increases across different stages of the interaction. While \citet[][]{Faria2025} found that $\mathrm{SFE_{H}}$ is the main driver of the improvement of sSFR, our findings differ during the merging stage, where the fuel (molecular gas) plays a more significant role in the increase of sSFR. Differences in the definition of gas fraction and efficiency, total gas \citep[][]{Faria2025} versus molecular gas, as well as differences in spatial and temporal resolution (continuous time versus discrete stages and radial profiles).
We also emphasize that, since our study is observational, we can only analyse the molecular gas that currently exists in each galaxy, which does not allow us to determine the origin of that gas, i.e., whether it comes from the interstellar medium of the progenitor galaxies or from the circumgalactic medium, as described in \citet[][]{Sparre2022}.
Furthermore, in this work, we do not trace the transformations of the gas through its different interstellar medium phases that lead to the formation of molecular gas, as explored in \citet{Moreno_2019}.
Also, we acknowledge that the effective radius is not a fixed property, but rather varies during mergers, as highlighted by simulation studies \citep[e.g.,][]{Moreno2021, Mercado_2025}. Since observations provide only a present-day snapshot, we cannot trace how the radius of a single galaxy changes throughout the merger process. Therefore, according to our classification scheme, we normalize all galaxies by their observed effective radius, allowing a consistent comparison of their properties across different merger stages.
These considerations highlight the complementarity between simulations and observations. While cosmological simulations provide a continuous view of the temporal evolution, allowing the track of gas across its different phases and origins, observational studies like ours directly constrain the molecular component currently available for star formation in galaxies. These two approaches will be key to understanding the complex interplay between fuel and efficiency in the variations of star formation during mergers. Also, to establish a more complete picture of how galactic interactions drive galaxy evolution.

Finally, it is worth highlighting the importance of studying star formation in extreme environments such as the Central Molecular Zone of the Milky Way, where a high concentration of molecular gas is inefficient for star formation \citep[e.g.,][]{Morris1996,Immer2012,Longmore2013,Barnes2017,Henshaw2022}. As we mentioned in \citet{Garay-Solis2023}, a deeper understanding of star formation behaviour in this region can help extrapolate these findings to the galactic scale, providing valuable insights into the impact of mergers on the evolution of galaxies.

\section{Summary and conclusions} 
\label{sec:conclusions}

In this study, we explore how the radial distributions of some internal properties of galaxies at different merger stages vary in comparison with the isolated sample. These properties are sSFR, SFE, and \fg, traced by the residuals \Dsfms, \Dsk, and \Dmgms, respectively, as well as the main driver of the enhanced and suppressed sSFR.
We use galaxies from the EDGE-CALIFA survey, which provides spatially resolved optical IFU and CO spectroscopy data. From the entire sample, we select galaxies with spaxels classified as star-forming regions and $\mathrm{\Sigma_{mol}\geq13~M_{\odot}~pc^{-2}}$ (see Section~\ref{subsec:Main_characteristics}).
Our sample includes 33 isolated galaxies and 34 galaxies in four merger stages.
We derive the best fits of spatially resolved scaling relations for the isolated sample: the rSFMS, the rSK relation, and the rMGMS. Based on these fits, we derive the residuals of the entire sample.
To account for the possibility that the correlations between residuals arise from statistical effects, we generate the radial distribution of mock data using the best fits of the aforementioned scaling relations and the fits and characterizations by \citet[][]{Barrera2023}.
We analyse the behaviour of \Dsk\ versus \Dmgms\ across different interaction stages compared to the control sample. Additionally, we investigate the main drivers of star formation (enhancement or suppression) at different radial bins. Finally, we compare the radial profiles of the residuals for each merger stage with those of the isolated sample.
Below, we summarize the main results of our analysis:

\begin{itemize}
    \item Pre-merger galaxies do not exhibit a significant increase in \fg, however, they show an increase in sSFR driven mainly by efficiency ($R<1.9~\Reff$).
    
    \item The merging sample shows increases in \fg\ ($R<1.9~\Reff$), but the most significant increase in sSFR is driven by efficiency ($R<0.5~\Reff$). The decrease in sSFR is efficiency-driven ($0.5<R<1.0~\Reff$), and the increase in sSFR is fuel-driven ($1.0<R<1.9~\Reff$).
    
    \item The post-merger sample shows an increase in \fg\ ($R<1.9~\Reff$) but exhibits a decrease in sSFR driven by efficiency ($R<1.2~\Reff$). 
    
    \item Galaxies in the merger remnant stage exhibit increases in the \fg\ ($R<0.5~\Reff$, $1.2<R<1.9~\Reff$), but they show a decrease in sSFR driven by fuel/efficiency ($R<1.2~\Reff$), and they show an increase in sSFR driven by efficiency ($1.2<R<1.9~\Reff$).

\end{itemize}   

In general, the merger process in galaxies increases the amount of molecular gas, though its distribution is highly non-uniform. Despite this, efficiency is the main driver of the variation (increase or decrease) of sSFR. Except for the merging stage, where the increase in the sSFR is driven primarily by fuel. Therefore, the main driver of star formation varies between radii and merger stages. This behaviour underscores the multifaceted impact that galaxy interactions and mergers have on star formation.
Therefore, this study highlights the importance of further exploring star formation in extreme environments. This is in order to understand the physical mechanisms that prevent significant increases in star formation despite the presence of molecular gas in galaxies at different merger stages.

\section*{Acknowledgements}

This work was funded in part by NSF AAG grants 2307440 to the University of Illinois and 2307441 to the University of Maryland.

Y.G.S. and J.B-B. acknowledge support from grant AG-101025 (DGAPA-PAPIIT, UNAM).

J.B-B acknowledges support from the DGAPA-PASPA 2025 fellowship (UNAM).

D.C. gratefully acknowledges the Collaborative Research Center 1601 (SFB 1601 sub-project B3) funded by the Deutsche Forschungsgemeinschaft (DFG, German Research Foundation) – 500700252. 

V.V. acknowledges support from the ANID BASAL project FB210003 and from ANID - MILENIO - NCN2024\_112.

A.Z.L.A. gratefully acknowledges the support provided by the Postdoctoral Program (POSDOC) of UNAM (Universidad Nacional Autónoma de México).

LC thanks the support by SECIHTI CBF-2025-I-2048 project.

\section*{Data Availability}

This study uses data from the publicly available EDGE-CALIFA Survey (https://doi.org/10.5281/zenodo.10256732). A detailed description is provided in \cite{Wong2024}. 



\bibliographystyle{mnras}
\bibliography{example} 



\appendix
\section{The main driver of the enhanced and suppressed star formation in bins of $\mathbf{\Delta SFMS}$}
\label{Appendix_fractions}

\begin{figure*}
\centering
    \includegraphics[width=1.0\linewidth]{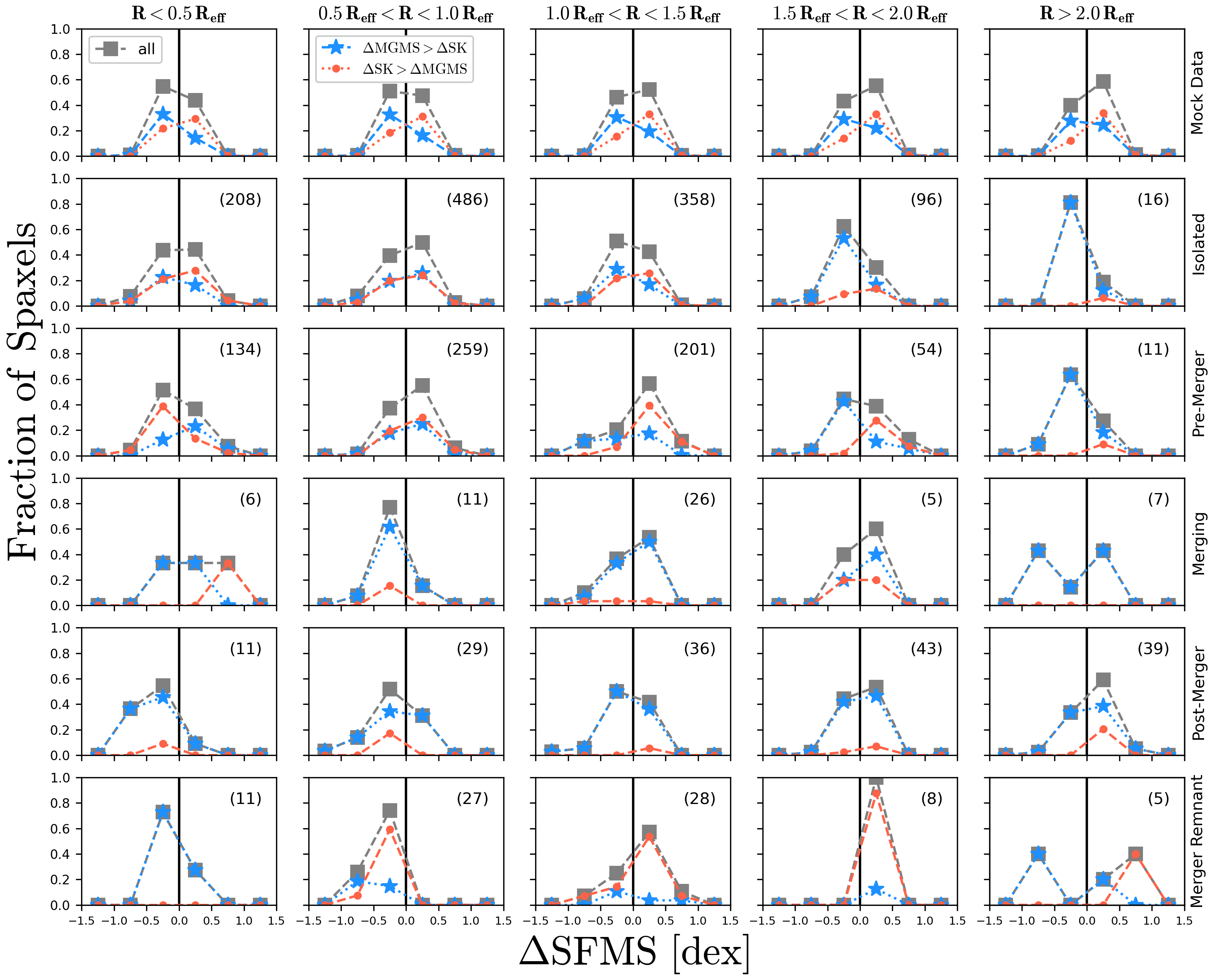}
    \caption{Fraction of star-forming regions explored in bins of \Dsfms\ for each merger stage, divided in bins of $0.5~\Reff$. Grey squares indicate the total number of regions in each bin. Blue stars indicate the regions that have a larger excess in their \fg\, than their SFE compared to the average population (sky-blue area in Fig.~\ref{fig:Stages_radio}). Red circles represent the regions with a larger excess in their SFE than their \fg\ (light-red area in Fig.~\ref{fig:Stages_radio}). The number of spaxels per radial bin is shown in parentheses in the top-right corner of each panel.
}
    \label{fig:Stages_driver_bines}
\end{figure*}

We explore the driver behaviour of both the enhanced and suppressed sSFR at each radio. To achieve this, we study the fraction of regions within \Dsfms\ bins across various radial bins.
In Fig.~\ref{fig:Stages_driver_bines}, we present from left to right the panels organised into bins of $0.5~\Reff$; from top to bottom, the panels show the mock data sample, the isolated sample, and the four merger stages, respectively. 
We present the total fraction of regions in \Dsfms\ bins (grey squares) for each merger stage and the mock data. For each bin, we indicate the fraction of regions in group (1) ($\Dmgms > \Dsk$, blue stars) and group (2) ($\Dsk > \Dmgms$, red circles), see Section~\ref{sec:Main_Drivers}.

In mock data, we find that both increases and decreases in sSFR ($\Dsfms>0$ and $\Dsfms<0$, respectively) are efficiency-driven across all the radial bins. However, a significant fraction of regions shows a decrease in the sSFR ($\Dsfms < 0$) at $R<0.5~\Reff$, while at $R>1.5~\Reff$, a larger fraction of regions shows an increase in the sSFR ($\Dsfms > 0$).
In the isolated sample at $R<0.5~\Reff$, we find that half of the regions exhibit an increase in sSFR. The enhanced sSFR appears to be efficiency-driven, whereas the main driver of the diminished sSFR is unclear, as the fraction belonging to the group (1) is similar to that belonging to the group (2). 
Also, at radii of 0.5 to $1.0~\Reff$, the main driver of star formation activity (increase/decrease) is not evident.  
This implies that neither molecular gas quantity nor efficiency is the dominant factor in these regions, but rather a combination of these might be driving the observed behaviour.
We find that the main driver of excess and deficit of the star formation is the efficiency for $1.0<R<1.5~\Reff$.
At $R>1.5~\Reff$, for the diminished sSFR, the primary driver is efficiency. 
For the enhanced sSFR, the main driver is not obvious.
Furthermore, we show that in the outer radii a larger fraction of regions exhibit decreased sSFR, $\sim70\%$ and $\sim81\%$ for $1.5<R<2.0~\Reff$ and $R>2.0~\Reff$, respectively.

In the pre-merger sample, we find that, compared to the isolated sample, the diminished/enhanced sSFR at $R<0.5~\Reff$ is fuel-driven. Furthermore, most of the regions in these radii exhibit a decrease in sSFR ($\sim56\%$).
At radii of 0.5 to $1.0~\Reff$, the behaviour of sSFR and its main drivers are similar to that of the isolated sample. As in the isolated sample in $1.0<R<1.5~\Reff$, the main driver of the enhanced sSFR ($\sim68\%$) is efficiency, however, for decreased sSFR it is unclear.
At $1.5 < R < 2.0~\Reff$, efficiency is the main driver of sSFR enhancement and diminishment.
The enhanced sSFR observed at $R > 2.0~\Reff$ is fuel-driven, while the main driver of diminished sSFR is efficiency.
In the central regions $(R<1.0~\Reff$) exhibit an increase in their sSFR due to both efficiency and fuel, the intermediate and outer regions $(R>1.0~\Reff$) show an increase in their star formation mainly driven by efficiency.

In the merging sample, we find behaviours different from those of the isolated sample.
We find that $\sim67\%$ of the regions exhibit an increase in sSFR, at $R<0.5~\Reff$, with half of these being fuel-driven for $0 < \Dsfms < 0.5$, and for $0.5 < \Dsfms < 1.0$ the increase is driven by efficiency. Also, we find that the diminished sSFR ($\sim33\%$) is efficiency-driven.
For $R>0.5~\Reff$, we find that the main driver of sSFR enhancement is fuel, whereas the diminished sSFR is mainly driven by efficiency, except in $1.5 < R < 2.0~\Reff$, where the main driver is unclear. In general, the enhanced star formation is driven by the amount of molecular gas. However, we find that although there is an increase in molecular gas, it is the efficiency that determines decreased star formation. 
In the post-merger sample, similar to the merging sample, we also find that the diminished sSFR is driven by efficiency across all radii. For $R>0.5~\Reff$, we find that enhanced sSFR is fuel-driven, while for $R<0.5~\Reff$, the main driver of sSFR enhancement remains unclear.
In the merger remnant sample, we find that, unlike the isolated sample, at $R<0.5~\Reff$, the enhanced sSFR is mainly fuel-driven, and the diminished sSFR is efficiency-driven. 
In contrast, at $0.5<R<1.0~\Reff$ the diminished sSFR is mainly fuel-driven, and the enhanced sSFR is not clear because in these radii, $100\%$ of regions exhibit a decrease in sSFR.
At $R>1.0~\Reff$, the sSFR enhancement is efficiency-driven; also, the sSFR diminishment is efficiency-driven for $R>2.0~\Reff$,  and in the remaining radii, the main driver is uncertain. 
Our results indicate that star formation efficiency is a key factor in both enhancing and suppressing star formation, regardless of the molecular gas content.
It is important to note that performing the same analysis, using regions within the coverage of $\sim 1~\Reff$ for each stage, in general, we find results consistent with those obtained by \citet{Garay-Solis2023}, who used integrated data in the central region ($\sim 1~\Reff$).


\bsp	
\label{lastpage}
\end{document}